\documentstyle[psfig,pre,multicol,aps]{revtex}

\def\sech{\mbox{sech}}

\begin{document}
\title{Solitons in anharmonic chains \\
with ultra-long-range interatomic interactions}

\author{Serge F. Mingaleev and Yuri B. Gaididei}
\address{Bogolyubov Institute for Theoretical Physics,
03143 Kiev, Ukraine}

\author{Franz G. Mertens}
\address{Physikalisches Institut, Universit\"at
Bayreuth, D-95440 Bayreuth, Germany}
\date{\today}
\maketitle
\begin{abstract}
We study the influence of long-range interatomic
interactions on the properties of supersonic pulse
solitons in anharmonic chains. We show that in the
case of {\em ultra-long-range} (e.g., screened Coulomb)
interactions {\em three different types} of pulse solitons
coexist in a certain velocity interval: one type is unstable
but the two others are stable. The high-energy stable
soliton is broad and can be described in the
quasicontinuum approximation. But the low-energy stable
soliton consists of two components, short-range and long-range
ones, and can be considered as a bound state of these
components.
\end{abstract}
\begin{multicols}{2}
\narrowtext

As is well known
\cite{Toda:1989:TNL,Flytzanis:1985:JPC},
anharmonic chains with interactions
between {\em nearest neighbors} can bear pulse solitons,
compressive localized excitations
which are very robust and propagate with supersonic
velocities without energy loss. Because of their coherence, the
solitons play an important role in determination of dynamical,
thermodynamic, and transport properties of one-dimensional
anharmonic systems \cite{Toda:1989:TNL}.
Among other things, they have been invoked in order to explain
energy transport in DNA \cite{Muto:1990:PRA}.

However, the interatomic interactions in {\em real systems}
are strictly speaking {\em long-ranged}. In particular, the
DNA molecule contains charged groups with
Coulomb interactions between them
\cite{Mingaleev:1999:JBP}.  Therefore, it is essential to
clarify how the long-range interactions (LRI's) can affect the
soliton features. It is generally believed that such interactions
are very small (in comparison with the anharmonic
interactions between nearest neighbors) and can be safely
neglected.  However, as we show in the present paper,
even very weak LRI's cause new qualitative effects
if the interactions are {\em ultra-long-ranged}.
A striking illustration is a chain with pure
(not screened) Coulomb interactions between charged
particles where the {\em sound velocity is infinite
regardless of the intensity} of these interactions.  As a
consequence the pulse solitons merely do not exist in such
a model (whereas the pure Coulomb interactions do not
prevent \cite{Bonart:1997:PLA} the existence
of immobile intrinsic localized modes therein).  Generally,
arbitrary LRI's introduce into the system a {\em new length
scale}, the so-called radius of the LRI's. If the radius of
the LRI's far exceeds the interatomic distance, the
competition between the length scales manifests itself in a
number of qualitative effects (see Refs.
\cite{Neuper:1994:PLA,Gaididei:1995:PRL}
for the exponential-law LRI's and Refs.
\cite{Ishimori:1982:PTP,Mingaleev:1998:PRE,Flach:1998:PD}
for the power-law LRI's).
The greater is the radius of the LRI's, the more
pronounced are these effects.

In this paper we show that {\em two types of stable pulse solitons
can coexist} in a certain interval of velocities in
anharmonic chains with {\em ultra-long-range} interatomic
interactions even if they are very weak.

Let us consider a chain of equally spaced particles of unit mass
whose displacements from equilibrium are $u_n(t)$ and the
equilibrium spacings are unity. The Hamiltonian of the system
is given by
\begin{eqnarray}
\label{sys:hamil}
H= \sum_n \biggl\{ \frac{1}{2} \biggl( \frac{du_n}{dt} \biggr)^2
+ V(u_{n+1}-u_n) \nonumber \\
+ \frac{1}{2} \sum_{m>n} J_{m,n} (u_m-u_n)^2 \biggr\} \; ,
\end{eqnarray}
with the anharmonic interactions
$V(w)= w^2/2-w^3/3$
between nearest neighbors and the harmonic LRI's
$J_{m,n}= J (e^{\alpha}-1) \, e^{-\alpha |m-n|}
/|m-n|^{s}$
between all particles of the chain.
Here $J$ characterizes the intensity of the LRI's whereas
$\alpha$ and $s$ determine their inverse radius.
The parameters $\alpha$ and $s$ are introduced to cover
different physical situations from the limit of
nearest-neighbor interactions
($\alpha \gg 1$ or $s \gg 1$) to the limit of ultra-long-range
interactions ($\alpha \ll 1$ and $s \leq 3$).
The Hamiltonian (\ref{sys:hamil}) generates
equations of motion of the form
\begin{eqnarray}
\label{sys:eq-wn}
\frac{d^2 w_n}{dt^2} &+& 2F(w_{n}) - F(w_{n+1})
- F(w_{n-1}) \nonumber \\
&+& \sum_{m \neq n} J_{m,n} (w_n-w_m)=0 \; ,
\end{eqnarray}
where $w_n=u_{n+1}-u_n$ are relative displacements and
$F(w) \equiv d V(w)/dw = w-w^2$ .

We assume in what follows that $\alpha \neq 0$ (the case
$\alpha=0$ has already been considered in Ref.
\cite{Mingaleev:1998:PRE});
in doing so we studied most extensively two cases: the
physically important screened Coulomb interactions ($s=3$)
and the Kac-Baker LRI's ($s=0$).
However, in view of the fact (tested numerically)
that all cases with $0 \leq s \leq 3$
lead to qualitatively the same results but the case $s=0$
allows also analytical consideration,
we discuss only the case $s=0$ from this point on.

In the quasicontinuum limit, treating $n$ as a continuous
variable [$n \to x$ , $w_n(t) \to w(x,t)$,
$w_m(t) \to e^{(m-n)\partial_x} w(x,t)$] and
keeping formally all terms in the Taylor expansion
of $e^{(m-n)\partial_x}$, the equation of motion 
(\ref{sys:eq-wn}) for $s=0$ can be cast in the operator 
form \cite{Gaididei:1995:PRL}
\begin{equation}
\label{sys:eq-wx}
[\partial^2_t - J Q(\alpha, \partial_x)] w(x,t)
- 4 \sinh^2 \biggl( \frac{\partial_x}{2} \biggr)
F(w)=0 \; ,
\end{equation}
where
\begin{equation}
\label{sys:Q}
Q(\alpha, \partial_x)= (e^{\alpha}+1)
\frac{4 \sinh^2(\partial_x/2)}
{\kappa^2 - 4 \sinh^2(\partial_x/2)}
\end{equation}
with $\kappa=2 \sinh(\alpha/2)$
is a linear pseudo-differential operator.
The speed of sound $c$ (which is an upper limit of
the group velocity of linear waves), determined by the
expression $c^2=1 + J (1+e^{-\alpha})/(1-e^{-\alpha})^2$ ,
grows indefinitely as $\alpha$ decreases.

\begin{figure}
\centerline{\hbox{
\psfig{figure=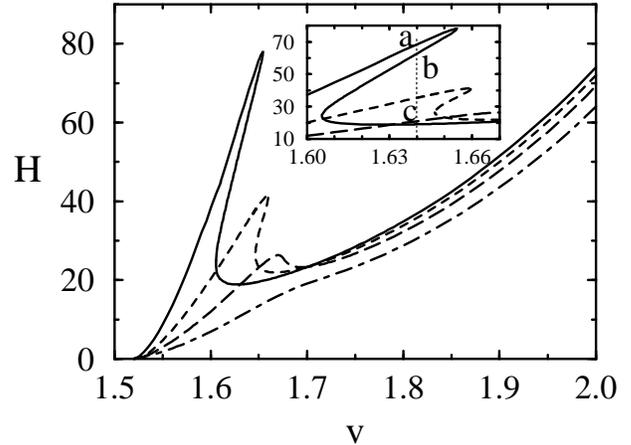,width=80mm,angle=0}}}
\caption{Energy of the pulse solitons versus velocity
found numerically for different
values of $\alpha$ and $J$ (the value of $J$ was chosen to
get constant $c=1.515$; see stars in Fig.
\protect\ref{fig:diagram}):
$\alpha=0.3$ and $J=0.05$ (dot-dashed line);
$\alpha=0.17$ and $J=0.0172$ (long-dashed line);
$\alpha=0.1$ and $J=0.0062$ (dashed line);
$\alpha=0.05$ and $J=0.0016$ (full line).}
\label{fig:energy}
\end{figure}

We are interested in the stationary soliton solutions
$w(x,t) \equiv w(x-vt)$
propagating with velocity $v$.
In this way we reduce our problem to a nonlinear eigenvalue
problem with $v$ being a spectral parameter.
Indeed, substituting $z=x-vt$ and using the continuum
approximation $4\sinh^2(\partial_x/2) \approx
\partial_x^2$, we can write Eq. (\ref{sys:eq-wx}) in
the form \cite{Gaididei:1995:PRL}
\begin{equation}
\label{sys:eq-wz}
(\partial_z^2-s_{+}^2) (\partial_z^2-s_{-}^2) w(z)
= \frac{12}{v^2} (\partial_z^2-\kappa^2) w^2(z) \; ,
\end{equation}
where the parameters $s_{\pm}$ are given by
\begin{eqnarray}
\label{sys:spm}
s^2_\pm = \frac{1}{2} && \bigg\{ \kappa^2 +
12 \frac{v^2-1}{v^2} \nonumber \\
\pm && \sqrt{\left( \kappa^2 - 12 \frac{v^2-1}{v^2}
\right)^2 + 48 \kappa^2 \frac{c^2-1}{v^2}}
\bigg\} \; .
\end{eqnarray}
The parameter $s_{+}$ is finite at all velocities
$v \geq c$ and tends to $\sqrt{12}$ for $v \to \infty$.
The parameter $s_{-}$ vanishes at $v=c$ and tends to
$\kappa$ for $v \to \infty$. Using the Green's function
method \cite{Mingaleev:1998:PRE} one can show that
stationary soliton solutions exist only for
supersonic velocities $v>c$.
The properties of these solitons are determined by the
ratio of $s_{+}$ and $s_{-}$.
In Fig. \ref{fig:energy} we plot the energy
of the soliton solutions of Eq. (\ref{sys:eq-wx})
which were found numerically using the method developed in
Ref. \cite{Mingaleev:1998:PRE}.

The soliton energy grows {\em monotonically} with the
velocity in the case of large $\alpha$ (see, e.g.,
$\alpha=0.3$ in Fig. \ref{fig:energy}).
In this case the soliton properties are qualitatively the
same as in the limit of nearest-neighbor interactions
(NNI's) for which Eq. (\ref{sys:eq-wx}) 
reduces to the Boussinesq equation. It is well known that
this equation has a 
sech-shaped soliton solution $w(z)= - 1.5 \, (v^2-c^2)/
\cosh^2 (\sigma z)$, where $\sigma=\sqrt{3(v^2-c^2)}$ is
the inverse soliton width. As indicated above, the 
energy $H \sim (v^2-c^2)^{3/2}$ of these solitons is 
monotonic function of
the velocity, which means that there is only one soliton
state for each given value of energy or velocity. 

\begin{figure}
\centerline{\hbox{
\psfig{figure=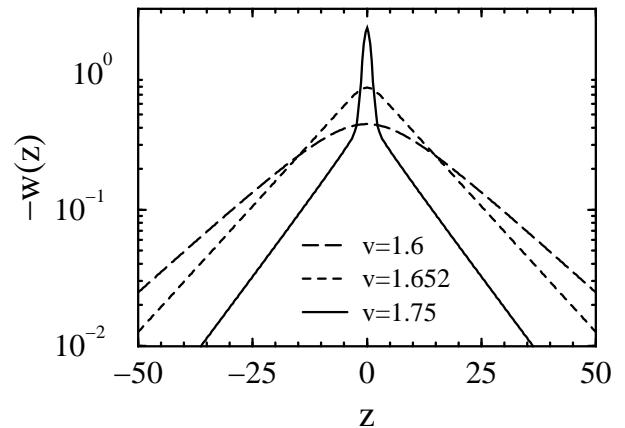,width=80mm,angle=0}}}
\caption{Shapes of pulse solitons
found numerically for different
velocities at $\alpha=0.17$ and $J=0.0172$.}
\label{fig:crest}
\end{figure}

In the case of small $\alpha$ the soliton properties
become much more interesting
\cite{Neuper:1994:PLA,Gaididei:1995:PRL}.
As it was recently shown
\cite{Gaididei:1995:PRL},
{\em two branches} of stable supersonic pulse solitons
should be distinguished in this case: low-velocity and
high-velocity solitons, separated by a gap with unstable
soliton states.

The solitons of the low-velocity branch are
broad (they have a width much larger than $1/s_{+}$),
and can be described by Eq. (\ref{sys:eq-wz})
in the approximation
$(\partial_z^2 - s_{+}^2) w \approx - s_{+}^2 w$.
In this approximation the soliton solutions
exist in a finite interval of velocities,
$c<v<v_{cr} \simeq \sqrt{(4c^2-1)/3} $, and change their shape
from the sech-form at $v \gtrsim c$ (see the case $v=1.6$
in Fig. \ref{fig:crest}) to the crest-form
$w(z) \sim \exp(-\alpha |z|/2)$ for $v \to v_{cr}$
(see the case $v=1.652$ in Fig. \ref{fig:crest}). Such
crest solitons (or {\em peakons}) were first introduced in the
theory of shallow water motion
\cite{Whitham:1974:LNW,Camassa:1993:PRL}.

The solitons of the high-velocity branch are made up 
of two components: $w=w_S(z)+w_L(z)$, where
the short-range component
\begin{equation}
\label{sys:ws}
w_S(z) \approx -\frac{s_{+}^2 v^2}{8} (1-2 \gamma)
\, \sech^2 \biggl( \sqrt{1-2 \gamma} \,
\frac{s_{+} z}{2} \biggr)
\end{equation}
is dominant in the center of the strain, while
the long-range component
$w_L(z) \simeq - \gamma \, (s_{+}^2 v^2/12)
\exp(-s_{-} |z|)$ is dominant in the tails
(see the case $v=1.75$ in Fig. \ref{fig:crest}).
It should be stressed that this division of the soliton 
body into two components is not just a mathematical trick. 
Our present numerical simulations testify that the solitons 
of the high-velocity branch can be considered as {\em bound
states} of the short-range and long-range components:  they
can be excited such that the relative distance between the
components oscillates.  However, such internal soliton
oscillations are highly damped and should not play an
important part in the nonlinear dynamics of the system.
The parameter $\gamma$ is determined by the equation 
\begin{equation}
\label{sys:gamma}
\gamma \sqrt{A\gamma^2 - 2(A+2)\gamma/3 + 1} =
3(A-1) \frac{s_{-}}{s_{+}} \sqrt{1-2\gamma} \; 
\end{equation}
with $A=\kappa^2/s_{-}^2$. This equation, derived in Ref. 
\cite{Gaididei:1995:PRL}, has been there analyzed for small 
values of $J$ and $\alpha$, for which it has a unique solution 
at all values of velocity $v$. 
It has been shown that the interplay of the components
$w_S(z)$ and $w_L(z)$ results in this case into {\em nonmonotonic}
dependence of soliton energy $H$ on the velocity 
(see, e.g., the case $\alpha=0.17$ in Fig. \ref{fig:energy}), so 
that there is an energy interval where three soliton
states with different velocities exist for each given value of
energy. As is shown in Ref. \cite{Gaididei:1995:PRL},
the low- and high-velocity states (with $dH/dv>0$)
are stable while the intermediate state (with $dH/dv<0$) is
unstable. 

\begin{figure}
\centerline{\hbox{
\psfig{figure=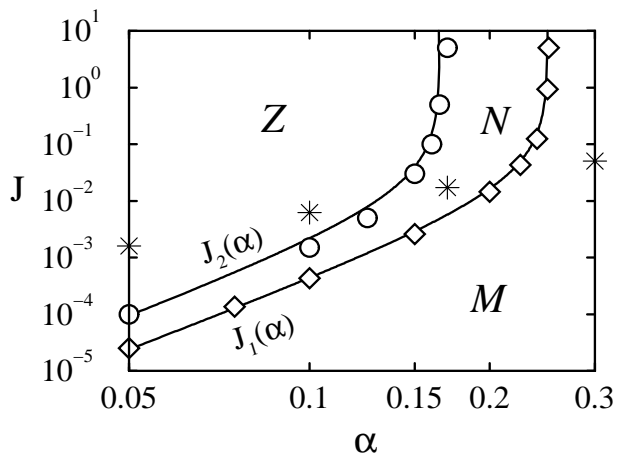,width=80mm,angle=0}}}
\caption{Three regions of the system parameters
with qualitatively different properties of
the pulse solitons. Full lines represent Eqs. (\ref{sys:J1}) 
and (\ref{sys:J2}). The points, marked off as circles and 
diamonds, were calculated numerically. 
Stars mark the parameters used in Fig. 
\protect\ref{fig:energy}.}
\label{fig:diagram}
\end{figure}

To sum up the foregoing, there is a demarcation line
$J_1(\alpha)$ which separates the plane $\{\alpha, J \}$ 
into two regions (see Fig. \ref{fig:diagram}), namely: 
the $M$-region (with a {\em monotonic} dependence of soliton energy on
the velocity) at $J<J_1(\alpha)$ and the $N$-region (with a 
{\em nonmonotonic} dependence of soliton energy on
the velocity) at $J>J_1(\alpha)$. Our numerical calculations 
(see Fig. \ref{fig:diagram}) validate the following estimation 
for $J_1(\alpha)$: 
\begin{equation}
\label{sys:J1}
J_1(\alpha) \simeq 0.23 \, \frac{\alpha^4}{\alpha_1^2 - 
\alpha^2} \; , 
\end{equation}
with $\alpha_1 \simeq 0.25$. 
The spectrum of stable soliton states is continuous and covers all
supersonic velocities in the $M$-region, while it has a gap (an
interval of velocities with unstable soliton states) in the 
$N$-region. Emerging at $J=J_1(\alpha)$ this gap increases initially
with growth of $J$. However, closer analytical examination of 
Eq. (\ref{sys:gamma}) shows that subsequently this gap starts to
decrease and disappears again at $J=J_2(\alpha)>J_1(\alpha)$, where 
\begin{equation}
\label{sys:J2}
J_2(\alpha) \simeq \frac{3}{8} \, \frac{\alpha^4}{\alpha_2^2 - 
\alpha^2} \; , 
\end{equation}
with $\alpha_2 \simeq 0.16$. Besides Eq. (\ref{sys:J2}) we have 
found analytical expressions for the soliton energy and impulse, 
all in a very good agreement with the numerical calculations. 
But due to lack of place we do not present these cumbersome 
formulas in the present short paper. Instead, we just discuss 
below the results obtained for $J>J_2(\alpha)$ with the intent 
to demonstrate that the soliton features in this region 
(lets call it $Z$-region) are qualitatively different from those 
in the $N$-region. 

\begin{figure}
\centerline{\hbox{
\psfig{figure=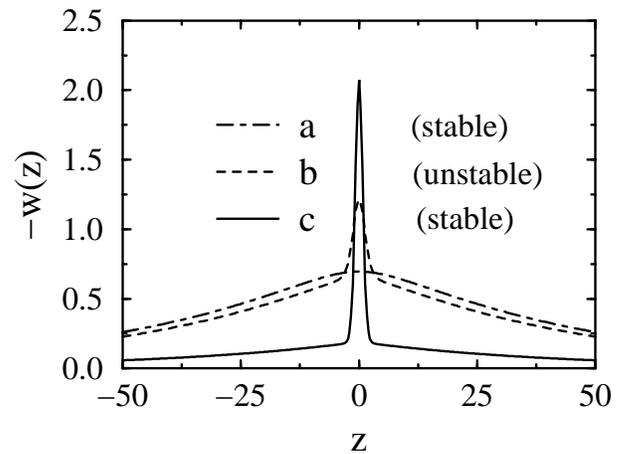,width=80mm,angle=0}}}
\caption{Shapes of pulse solitons
which coexist at the same velocity
$v=1.64$, where $\alpha=0.05$ and $J=0.0016$. These solitons
are also indicated on the inset of Fig.
\protect\ref{fig:energy}.}
\label{fig:form-v}
\end{figure}

\begin{figure}
\centerline{\hbox{
\psfig{figure=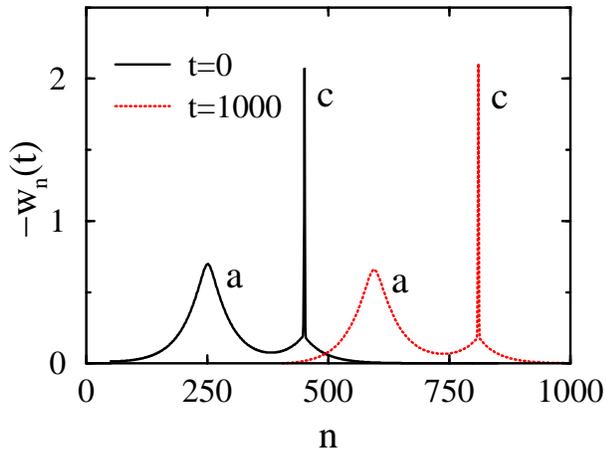,width=80mm,angle=0}}}
\caption{Demonstration of the dynamical stability
of the two stable pulse solitons ($a$ and $c$ in Fig.
\protect\ref{fig:form-v}) which propagate with
the same velocity $v=1.64$. Here
$\alpha=0.05$ and $J=0.0016$.}
\label{fig:dynamic}
\end{figure}

Indeed, when $J$ exceeds $J_2(\alpha)$ there appears an interval of 
velocities in which Eq. (\ref{sys:gamma}) has two real solutions. 
They correspond to two different types of two-component pulse 
solitons which coexist at the same velocity 
($b$ and $c$ in Fig. \ref{fig:form-v}). Accordingly, the 
dependence of the soliton energy on the velocity for 
$J>J_2(\alpha)$ is not merely nonmonotonic but takes
on a $Z$-shaped {\em multivalued} form (see, e.g., the cases 
$\alpha=0.1$ and $\alpha=0.05$ in Fig. \ref{fig:energy}). 
The possibility of such a dependence has been predicted in 
Ref. \cite{Neuper:1994:PLA} using a variational approach.
At that time, however, this prediction was met with
disbelief and considered as an artifact of
variational approach. But as we prove numerically in the 
present paper, the $Z$-region really exists. In this region
there is an interval of velocities where three soliton
states of quite different shapes (see Fig. \ref{fig:form-v})
and energies {\em coexist at the same velocity}.
The soliton state with intermediate energy
($b$ in Fig. \ref{fig:form-v}) is always unstable.
But the high-energy and low-energy solitons ($a$ and $c$ in Fig.
\ref{fig:form-v}) are usually (when $dH/dv>0$) stable.
The high-energy soliton on the low-velocity
branch is broad and has a single component. But the
low-energy soliton on the high-velocity
branch and the soliton state with intermediate energy both
consist of two components, short-range and long-range ones.
The coexistence of two different types of stable pulse solitons at
the same velocity causes new interesting phenomena, e.g.,
the synchronous propagation of two solitons with quite different
widths (see Fig. \ref{fig:dynamic}).

In conclusion, we show that the properties of pulse solitons 
in anharmonic chains with the long-range interatomic interactions 
are conveniently mapped onto the $\alpha J$-plane 
(see Fig. \ref{fig:diagram}). One can recognize in this plane 
three regions with qualitatively different properties of the pulse
solitons. The $M$- and $N$-regions were distinguished and discussed
in Refs. \cite{Neuper:1994:PLA,Gaididei:1995:PRL} 
whereas the $Z$-region (that is the region of {\em ultra-long-range} 
interatomic interactions) is proven to exist in the present 
paper. In this region there exists an interval of velocities where 
two types of stable pulse solitons coexist at each value of the 
velocity. The high-energy soliton is broad and has only a single
component whereas the low-energy soliton consists of two components,
short-range and long-range ones, and can be considered as a
bound state of these components. 
It should be stressed that this phenomenon occurs even for
LRI's of {\em very small intensity} (hundreds times
less than the intensity of NNI's, as is seen on Fig.
\ref{fig:diagram}) if only the radius of LRI's is large
enough. It is important that the coexistence of two types of
stable solitons does occur not only for the 
Kac-Baker LRI's discussed in this paper; on the 
contrary, it is rather a common phenomenon. In particular, 
we also have shown that it exists in anharmonic chains
with weakly screened Coulomb interactions between charged
particles.


Two of us (Yu.G.\ and S.M.) thank the University of Bayreuth,
where the main part of this work was done, for the 
hospitality.  We also acknowledge the support provided by
the DLR project UKR--002--99 of the scientific and 
technological cooperation between Germany and Ukraine.


\begin{thebibliography}{10}

\bibitem{Toda:1989:TNL}
M. Toda, {\em Theory of nonlinear lattices}, Vol.~20 of
{\em Springer Series in Solid State Sciences}, 2 ed.
(Springer, Berlin, 1989).

\bibitem{Flytzanis:1985:JPC}
N. Flytzanis, S. Pnevmatikos, and M. Remoissenet,
J. Phys. C {\bf 18}, 4603 (1985).

\bibitem{Muto:1990:PRA}
V. Muto, P.~S. Lomdahl, and P.~L. Christiansen,
Phys. Rev. A {\bf 42}, 7452 (1990).

\bibitem{Mingaleev:1999:JBP}
S.~F. Mingaleev, P.~L. Christiansen, Yu.~B. Gaididei, 
M. Johansson, and K. {\O}. Rasmussen, 
J. Biol. Phys. {\bf 25}, 41 (1999).

\bibitem{Bonart:1997:PLA}
D. Bonart, Phys. Lett. A {\bf 231}, 201 (1997); 
D. Bonart, T. R{\"o}ssler, and J.~B. Page,
Physica D {\bf 113}, 123 (1998).

\bibitem{Neuper:1994:PLA}
A. Neuper, Yu. Gaididei, N. Flytzanis, and F. Mertens,
Phys. Lett. A {\bf 190}, 165 (1994).

\bibitem{Gaididei:1995:PRL}
Yu. Gaididei, N. Flytzanis, A. Neuper, and F.~G. Mertens,
Phys. Rev. Lett. {\bf 75}, 2240 (1995); 
Physica D {\bf 107}, 83 (1997).

\bibitem{Ishimori:1982:PTP}
Y. Ishimori, Prog. Theor. Phys. {\bf 68}, 402 (1982).

\bibitem{Mingaleev:1998:PRE}
S.~F. Mingaleev, Yu.~B. Gaididei, and F.~G. Mertens,
Phys. Rev. E {\bf 58}, 3833 (1998).

\bibitem{Flach:1998:PD}
S. Flach, Physica D {\bf 113}, 184 (1998); 
Phys. Rev. E {\bf 58}, R4116 (1998).

\bibitem{Whitham:1974:LNW}
G.~B. Whitham, {\em Linear and nonlinear waves}
(Wiley, New York, 1974).

\bibitem{Camassa:1993:PRL}
R. Camassa and D.~D. Holm,
Phys. Rev. Lett. {\bf 71}, 1661 (1993).

\end{thebibliography}


\end{multicols}
\end{document}